\documentclass[10pt,a4paper]{article}
\usepackage[utf8]{inputenc}
\usepackage{float}
\usepackage{graphicx}
\usepackage[left=2cm,right=2cm,top=2cm,bottom=2cm]{geometry}
\usepackage{authblk}
\begin{document}

\title{Permeability prediction of organic shale with generalized lattice Boltzmann model considering surface diffusion effect}
\author[a,b]{Junjian Wang}
\author[c,b]{Li Chen}
\author[b]{Qinjun Kang
\thanks{Corresponding author at: EES-16, LANL, Los Alamos, NM, 87545 Email: qkang@lanl.gov}}
\author[a]{Sheik S Rahman}
\affil[a]{School of Petroleum Engineering, University of New South Wales, Sydney, NSW, Australia, 2033}
\affil[b]{Earth and Environmental Sciences Division, Los Alamos National Laboratory, Los Alamos, NM, USA, 87545}
\affil[c]{Key Laboratory of Themo-Fluid Science and Enginnering of MOE, School of Energy and Power Engineering, Xi'an Jiaotong University, Xi'an, Shanxi, China, 710049}

\date{}
\maketitle

\abstract{Gas flow in shale is associated with both organic matter (OM) and inorganic matter (IOM) which contain nanopores ranging in size from a few to hundreds of nanometers. In addition to the noncontinuum effect which leads to an apparent permeability of gas higher than the intrinsic permeability, the surface diffusion of adsorbed gas in organic pores also can influence the apparent permeability through its own transport mechanism. In this study, a generalized lattice Boltzmann model (GLBM) is employed for gas flow through the reconstructed shale matrix consisting of OM and IOM. The Expectation-Maximization (EM) algorithm is used to assign the pore size distribution to each component, and the dusty gas model (DGM) and generalized Maxwell-Stefan model (GMS) are adopted to calculate the apparent permeability accounting for multiple transport mechanisms including viscous flow, Knudsen diffusion and surface diffusion. Effects of pore radius and pressure on permeability of both IOM and OM as well as effects of Langmuir parameters on OM are investigated. Moreover, the effect of total organic content and distribution on the apparent permeability of the reconstructed shale matrix is also studied. It is found that the distribution of OM and IOM has a negligible influence on apparent permeability, whereas the total organic content and the surface diffusion play a significant role in gas transport in shale matrix.}

\section{Introduction}

Shale gas reservoirs contain a significant proportion of hydrocarbon energy, and the successful exploitation of such resource plays an increasingly important role in meeting world's demand for natural gas. Organic shale is known to be fine grained sedimentary rocks consisting of inorganic matter (IOM) and organic matter (OM) with pore sizes ranging from nano- to meso- scale\cite{loucks2009morphology} \cite{sondergeld2010petrophysical}, in each component of which are involved different flow mechanisms\cite{javadpour2007nanoscale}. In the literature, gas flow on nano-to micro-scale in shale is often  referred to as rarefied gas flow\cite{singh2013nonempirical}, where the mean free path of gas is comparable to the characteristic length of the micropores or throats. Gas flow in shale matrix usually leads to a deviation from the continuum theory\cite{karniadakis2006microflows}. Moreover, previous studies have confirmed that the amount of adsorbed gas constitutes about $20\%$-$80\%$ of the total gas in place of shale gas reservoir\cite{wu2014generalised}, and the surface diffusion of adsorbed gas is an important transport mechanism in these reservoirs\cite{wu2014apparent}\cite{wu2015model}. A physics-based understanding of gas transport mechanism in shale including non-continuum behaviours and surface diffusion is essential for the development of accurate descriptive transport simulators to predict fluid flow and transport in shale.

For decades, the problem of modelling gas transport in narrow pores and confined spaces in shale has attracted considerable attention among petroleum engineers. Generally speaking, two approaches have been proposed to describe the gas transport and to calculate apparent permeability of organic shale. The first approach is to modify the non-slip boundaries in continuum model by accounting for slip boundary conditions. Beskok and Karniadaki\cite{beskok1999report} derived a unified Hagen-Poiseuille-type formula to take account of slip flow, transition flow and free molecular flow. Later, Civan\cite{civan2010effective} and Florence et al.\cite{florence2007improved} proposed different forms of rarefaction coefficient in Beskok-Karniadaki model. Xiong et al.\cite{xiong2012fully} introduced a capillary model by adding the mass transfer of adsorbed gas into Beskok-Karniadaki empirical equation to study the impact of the adsorbed gas and surface diffusion on gas apparent permeability. The second approach is the superposition of various transport mechanisms. Javadpour\cite{javadpour2007nanoscale} combined slip flow and Knudsen diffusion into gas flux equation and derived an equation for apparent permeability. Freeman et al.\cite{freeman2011numerical} applied dusty gas model (DGM) to account for Knudsen diffusion in shale gas reservoir. Singh et al.\cite{singh2013nonempirical} combined viscous flow with Knudsen diffusion in their non-empirical apparent permeability model(NAP), and validated the model with previous experimental data. The results show that the NAP can be used for Kn less than unity. Wu et al.\cite{wu2014apparent} further proposed two weighted factors for viscous  flow and Knudsen diffusion, respectively. The surface diffusion was also considered in their apparent permeability model. 

The limitation of the application of analytical or semi-analytical models to gas flow in porous media is that the pore structure is usually relatively simple, such as capillaries\cite{javadpour2007nanoscale}\cite{singh2013nonempirical}\cite{xiong2012fully}. Such simplification might produce erroneous results because the pore structures in shale are very complex, as detected by well-established characterization techniques such as SEM\cite{loucks2009morphology}\cite{curtis2010structural}. To improve this, different pore-scale models have been proposed  to link the micro-structure of the porous media with fluid flow characteristics. Among them, the lattice Boltzmann (LB) method has gone through significant improvements over the past years and has become a viable and efficient substitute for conventional N-S solvers in many flow problems especially porous flow and multiphase flow\cite{chen2014critical}. Because of its inherent kinetic nature, the LB method has attracted a huge interest in its extension to simulating micro-gaseous flows, and tremendous efforts have been made to advance the LBM since 2002\cite{nie2002lattice}\cite{guo2007discrete}\cite{chen2015nanoscale}. With the implementation of appropriate slip boundary conditions and/or effective relaxation time, the LBM was successfully extended for simulation of gaseous flows in slip flow and transition  flow regimes, and these LBM approaches have been applied to study gas transport in shale gas reservoir\cite{ren2015lattice}\cite{Fathi2013lattice}\cite{wang2015lattice}. However, because of the complexity of boundary conditions, most of the applications of the slip-based LBM are limited to single channel or bundle of channels\cite{wang2015lattice}. Recently, Chen et al.\cite{chen2015nanoscale}\cite{chen2015pore} proposed a LB model based on the Dusty gas model(DGM) to predict the apparent permeability of shales with complex porous structures, where the complexity of the slip boundary conditions are avoided. Very recently, Chen et al.\cite{chen2015generalized} improved the generalized lattice Boltzmann model (GLBM) proposed by Guo and Zhao\cite{guo2002lattice} for fluid flow through porous media by including the Klinkenberg effect, and performed several simulations based on heterogeneous shale matrix with natural fractures, organic matter and inorganic minerals\cite{chen2015generalized}\cite{chen2015permeability}.   

In this study, we present a novel adaptation of the GLBM  with slip effect proposed by Guo and Zhao\cite{guo2002lattice} and Chen et al.\cite{chen2015generalized} for microgas flow in porous shale with surface diffusion further considered. The novelty of the present study is the use of the DGM-Generalised Maxwell-Stefan(GMS) approach to calculate the local permeability taking the adsorbed gas and surface diffusion into account. The rest of the paper is as follows. The mathematical and numerical models for predicating apparent permeability of organic shale is introduced in Section 2, and the validation of numerical models is also shown in this section. In section 3, firstly the permeability of IOM and OM are discussed, and then the effects of component distribution and organic content on apparent permeability of reconstructed shale sample are analysed. Finally, some conclusions are drawn in section 4.  

\section{Model description}
\subsection{Dusty gas model and generalised Maxwell-Stefan model}
\begin{figure}[H]
\centering
\includegraphics[scale=0.6]{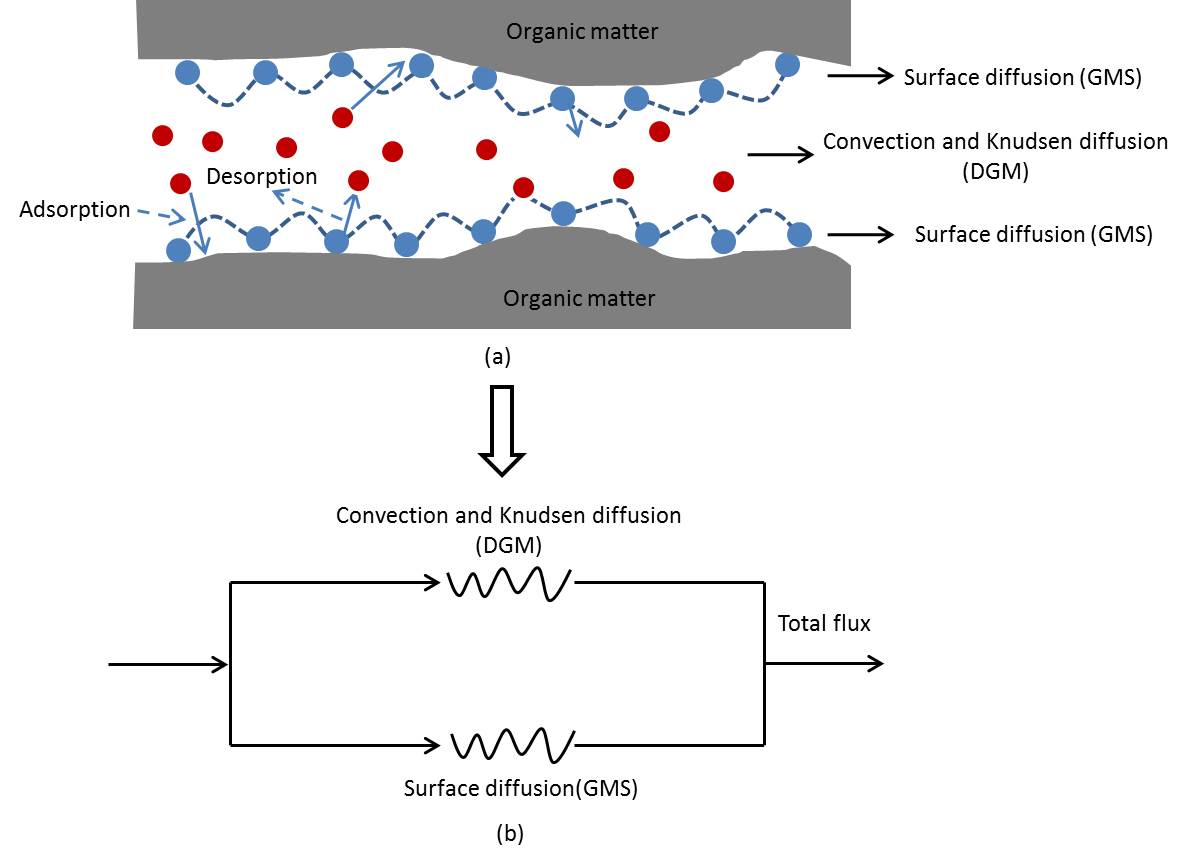}
\caption{Schematic diagram of gas transport in organic pores of shale matrix based on DGM-GMS model proposed by Krishna and co-workers\cite{krishna1993problems}\cite{krishna1997maxwell} (a) Gas transport mechanisms in organic pores (b) Conceptual models for different gas transport}
\label{figschematic}
\end{figure}

The transport of non-adsorbable gas through porous media is in general caused by concentration or pressure gradients. It has been confirmed experimentally and mathematically that the corresponding fluxes can be calculated with high accuracy using the DGM or Knudsen-like theory\cite{cunningham1980diffusion}. However, most recent studies indicate that the above applications have failed to model the gas transport in the presence of the adsorbed gas\cite{bhatia2011molecular}. To take account of the adsorbed gas effect, Krishna and co-workers\cite{krishna1993problems}\cite{krishna1997maxwell} extended the Maxwell-Stefan formulation in the spirit of the DGM by introducing a generalised Maxwell-Stefan model (GMS) for surface diffusion of adsorbed gas(See Fig.\ref{figschematic}). With the premise that the DGM appropriately describes the transport in the ``quasi-bulk" phase located in the pore space whereas the GMS describes gas molecular transport at the surface, the total molar flux is expressed by:

\begin{equation}
N_{total}= N_{DGM} + N_{GMS},
\end{equation}
where $N_{DGM}$ is the contribution to the flux from the DGM, in which adsorption and surface diffusion is ignored, and $N_{GMS}$ is the contribution from the GMS for surface diffusion. Detailed descriptions of DGM and GMS are given in the next two subsections, respectively. 

\subsubsection{Dusty gas model}
The DGM is based on the combination of the Maxwell-Stefan diffusion equations and the characteristics of mass transfer in porous media. The basic idea of DGM is to consider the solid as a dummy species of infinite mass, which is constrained by unspecified external forces and has zero drift velocity. For a single species $i$ in a n-component mixture the following flux equation of the DGM holds\cite{mason1983gas}:

\begin{equation}
\label{EqGDGM}
-\frac{P}{R T} \nabla x_i -\frac{x_i}{R T} (1+ \frac{K_0}{\eta D_{K,i}}P)\nabla P = \sum_{j=1,j \neq i}^n \frac{x_j N_i - x_i N_j}{\varepsilon / \tau D_{ij}^0} +\frac{N_i}{D_{K,i}},
\end{equation}
where $P$ is pressure, $R$ is the universal gas constant, $T$ is temperature, $x_i$ is the molar fraction of species $i$, $D_{ij}^0$ is the binary molecular diffusivity in gas phase, and $\varepsilon / \tau$ is the ratio of porosity to tortuosity. The Knudsen diffusivity $D_{k,i}$ of species $i$ is defined as:
\begin{equation}
D_{K,i}=\frac{4}{3}K_c\sqrt{\frac{8RT}{\pi M_i}},
\end{equation}
where $M_i$ is the Molecular weight of species $i$. In the case where the pore space is assumed to have a diameter $d_p$, the values of  Knudsen coefficient, $K_c$ and the permeability of the porous medium quantifying the viscous flux, $K_0$ can be related as\cite{tuchlenski1998experimental}:
\begin{equation}
K_c= \frac{8}{d_p}K_0=\frac{\varepsilon}{\tau}\frac{d_p}{4}.
\end{equation} 

If the system contains only one species, Eq.\ref{EqGDGM}  can be simplified as:
 
\begin{equation}
\label{EqDGM}
N_{DGM}=-\frac{1}{RT}(D_K +\frac{K_0}{\eta}P)\frac{\partial P}{\partial r}.
\end{equation} 

\subsubsection{Generalized Maxwell-Stefan Model}
The GMS model is based on the assumption that the movement of species is caused by a driving force balanced by the friction that the moving species experience both from each other and from their surroundings. The diffusion of adsorbed species satisfies\cite{krishna1997maxwell}:
\begin{equation}
\label{EqGMS}
-\frac{\theta_i(1-\varepsilon)}{RT}\nabla \mu_i = \sum^n_{j=1,j\neq i}\frac{\theta_j N_i^s-\theta_i N_j^s}{\rho_p q_{sat} D_{ij}^s} +\frac{N_i^s}{\rho_p q_{sat} D_i^s},
\end{equation}
where $\mu_i$ is the chemical potential of $i$, $\theta_i$ is the fractional coverage, $\rho_p$ is the density of particle or solid skeleton, $q_{sat}$ is the saturation surface concentration, $D_{ij}^s$ is the Maxwell-Stefan counter-sorption diffusivity, and $D_i^s$ is the Maxwell - Stefan diffusivity of species $i$. The surface fluxes $N_i^s$ of the diffusing adsorbed species are defined as:

\begin{equation}
N_i^s = (1-\varepsilon)q_{sat}\theta_i \textbf{u}_i,
\end{equation}
where $\textbf{u}_i$ is species velocity. Assuming equilibrium between the surface and the bulk phase, the following relationship for the surface chemical potential $\mu_i$ of species $i$ holds:
\begin{equation}
\mu_i = \mu_i^0 + RT ln(f_i),
\end{equation}
where $\mu_i^0$ is the chemical potential in the chosen standard state and $f_i$ is the fugacity of species $i$ in the bulk fluid mixture. For not too high system pressures, the component partial pressures, $P_i$, can be used instead of the component fugacity. Then the surface chemical potential gradients can be expressed in terms of the gradients of the surface occupancies by introduction of the matrix of thermodynamics factors:

\begin{equation}
\label{Eqmu}
\frac{\theta_i}{RT}\nabla \mu_i = \sum^n_{j=1} \Gamma_{ij}\nabla \theta_j,
\end{equation}
and the elements of thermodynamic matrix $\Gamma$ can be expressed as: 
\begin{equation}
\Gamma_{ij} \equiv \theta_i \frac{\partial ln P_i}{\partial \theta_j}.
\end{equation} 

Assuming that the Langmuir equation to be valid and the adsorption equilibrium to be established, the fractional coverage $\theta_i$ can be written in terms of the Langmuir parameters:
\begin{equation}
\label{EqLangmuir}
\theta_i=\frac{q_i}{q_{sat}}=\frac{b_iP_i}{1+\sum_{j=1}^nb_jP_j},
\end{equation}
where $b_i$ is the Langmuir constant, $q_{i} = q_{m,i}/(1-\varepsilon)$ is the solid volume dependent adsorbed phase concentration, and $q_{m,i}$ is the equilibrium mass dependent loading which is measured during the Langmuir adsorption experiment. 
 
Based on Eq.\ref{EqGMS} and Eq. \ref{Eqmu}, for a single component surface diffusion, the molar flux of surface diffusion is:
\begin{equation}
\label{EqGMS2}
N_{GMS} = (1-\varepsilon)q_{sat}\frac{D_s}{1-\theta}\nabla \theta,
\end{equation}
where $D_s$ is the Maxwell-Stefan surface diffusivity. And Langmuir equation (Eq. \ref{EqLangmuir}) can be simplified to:
\begin{equation}
\label{EqsimpleLangmuir}
\theta =\frac{q}{q_{sat}} = \frac{b P}{1+b P}.
\end{equation}

Based on Eq.\ref{EqsimpleLangmuir}, the gradient of the adsorbed phase concentration can be further expressed in terms of the partial pressure gradient as follows:
\begin{equation}
\label{EqpartialLangmuir}
\frac{\partial q}{\partial r} = q_{sat} \frac{b}{(1+bP)^2} \frac{ \partial p}{\partial r},
\end{equation}
where $r$ represents the pore radius. Substituting Eq.\ref{EqpartialLangmuir} to Eq. \ref{EqGMS2}, the surface diffusion flux in terms of partial gradient satisfies:
\begin{equation}
\label{EqGMS3}
N_{GMS} =  (1-\varepsilon)q_{sat}\frac{b D_s}{1+bP}\frac{\partial P}{\partial r}. 
\end{equation}

For a single gas species the combination of the DGM (Eq.\ref{EqDGM}) and the GSM (Eq.\ref{EqGMS3} ) results in the total flux:
\begin{equation}
\label{EqNtotal}
N_{total}=N_{DGM} + N_{GMS} = -(\frac{1}{RT}(D_K +\frac{K_0}{\eta}p) + (1-\varepsilon)q_{sat}\frac{b D_s}{1+bP})\frac{\partial P}{\partial r}, 
\end{equation}
and the apparent permeability $k_{app}$ satisfies:
\begin{equation}
\label{Eqm}
m_{total}= N_{total}M\pi(\frac{d_p^2}{4})=\rho_g \pi (\frac{d_p^2}{4}) \frac{K_{app}}{\eta}\frac{\partial p}{\partial x},
\end{equation}
where $m_{total}$ is the total mass flow rate, and $M$ is the molecular weight of Methane. As the surface diffusion of adsorbed gas only happens in organic matter, substituting Eq. \ref{EqDGM} into Eq. \ref{Eqm}, the permeability of inorganic matter is:
\begin{equation}
\label{Eqkappiom}
K_{app} = \frac{\eta M}{\rho_g}(\frac{1}{RT}(D_K +\frac{K_0}{\eta}p)),  
\end{equation}
and with Eq. \ref{EqNtotal} and Eq. \ref{Eqm}, the permeability of organic matter is:
\begin{equation}
\label{Eqkappom}
K_{app} = \frac{\eta M}{\rho_g}(\frac{1}{RT}(D_K +\frac{K_0}{\eta}p) + (1-\varepsilon)q_{sat}\frac{b D_s}{1+bP}).  
\end{equation}

\subsubsection{Validation}
Tuchlenski et al.\cite{tuchlenski1998experimental} reported the experimental data of several gases flowing through the porous Vycor glass membrane. The experimental data of Carbon dioxide flowing through the porous Vycor glass membrane under temperature 293K and 343K were used here to validate the formulation to calculate the apparent permeability of organic matter with surface diffusion considered(Eq.\ref{Eqkappom}). The structure properties and geometry of the Vycor membrane are shown in Table \ref{Tableproperties}. 

\begin{table}[H]
\centering
\caption{Structure properties of the Vycor membrane}
\label{Tableproperties}
\begin{tabular}{c|c}
\hline 
Inner membrane radius $r_1$ (mm) & 3.9   \\
\hline
Outer Membrane radius $r_2$ (mm)  &5  	\\
\hline
Membrane length $L$ (mm) 	& 100		\\
\hline
Apparent density $\rho_a$ ($g/cm^{3}$) & 1.472 \\
\hline
skeleton density $\rho_p$ ($g/cm^{3}$)  & 2.057 \\
\hline
Porosity   & 0.2842 \\
\hline
The ratio of porosity to tortuosity $\varepsilon/\tau$  				 &0.03 \\
\hline

\end{tabular}
\end{table}

The $CO_2$ adsorption isotherms were modelled by Langmuir equation, and the obtained parameters are given in Table \ref{Tablelangmuir}. 

\begin{table}[ht1]
\centering
\caption{Langmuir parameter of $CO_2$ }
\label{Tablelangmuir}
\begin{tabular}{c|c|c|c}
\hline 
T (K) & $D_s (m^2/s)$  & $q_{sat} (mol/ m^{3})$  & $b (P_a^{-1})$ \\
\hline

293K  & $2.2 \times 10^{-9}$ & $3135$ & $0.467\times 10^{-5}$ \\

343K  & $2.8 \times 10^{-9}$ & $2255$ & $0.191\times 10^{-5}$ \\

\hline

\end{tabular}
\end{table}

\begin{figure}[H]
\centering
\includegraphics[scale=0.26]{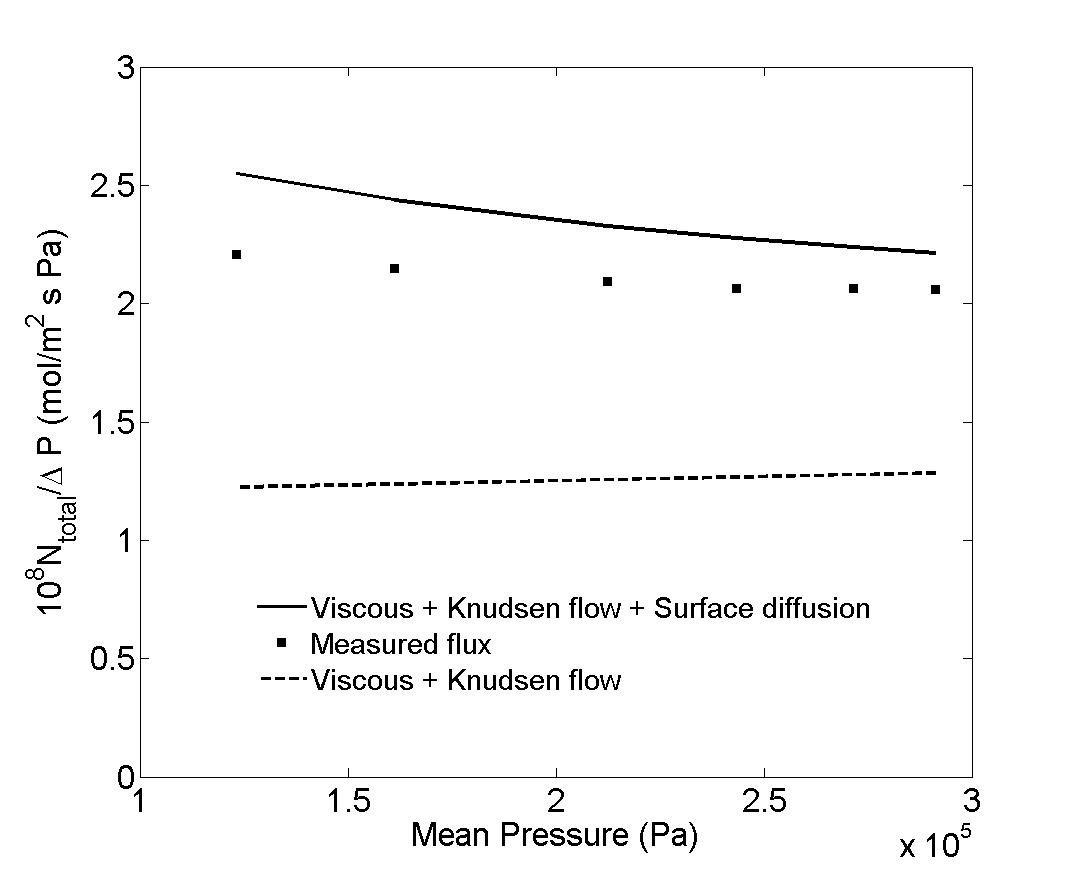}
\caption{Permeation of carbon dioxide through Vycor at 293 K}
\label{fig293}
\end{figure}

\begin{figure}[H]
\centering
\includegraphics[scale=0.26]{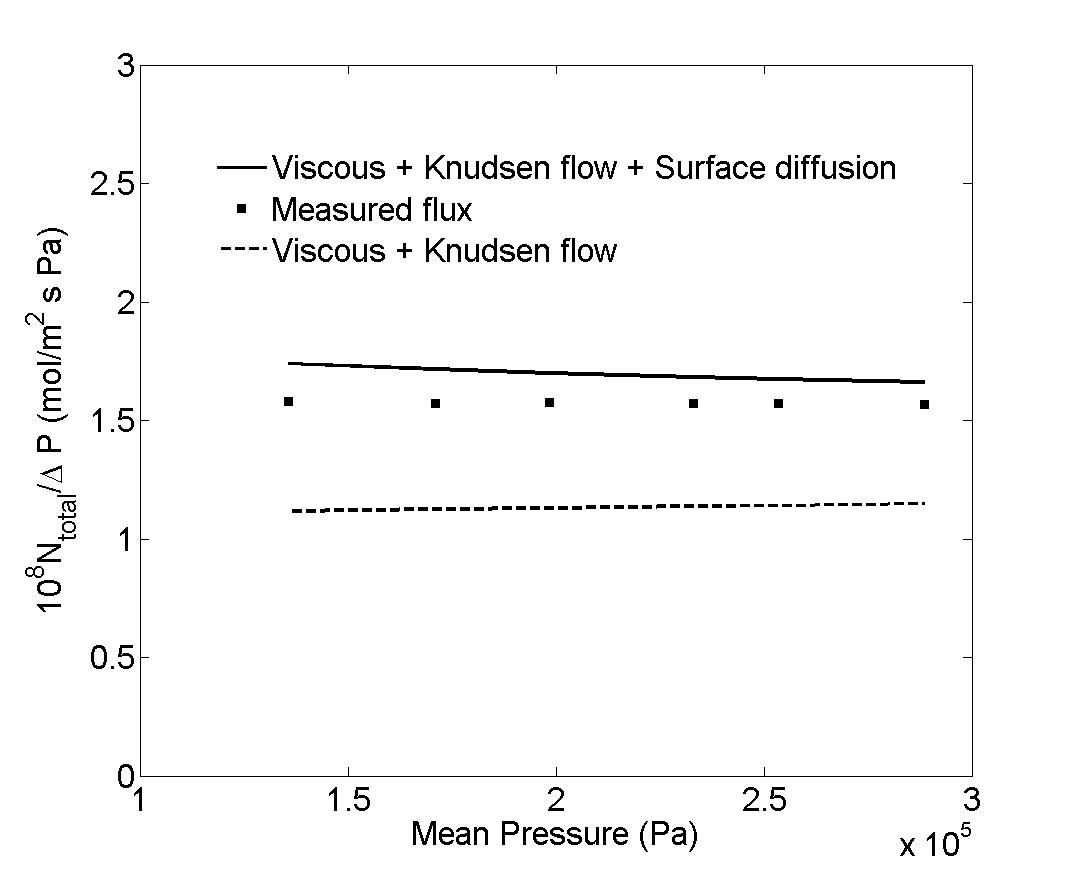}
\caption{Permeation of carbon dioxide through Vycor at 343 K}
\label{fig343}
\end{figure}

Experimental data and calculations of total molar flow rate, $N_{total}$ are given in Fig. \ref{fig293} and Fig. \ref{fig343}. The model results match reasonably well with experimental observations, with the deviations below $15\%$. As predicted by theory a slight decrease of the permeation data was observed for increasing pressure. It can also be concluded from Fig. \ref{fig293} and Fig. \ref{fig343} that for the given experimental conditions, the surface transport was only slightly smaller than the transport in the bulk phase, and the DGM without surface diffusion will lead to a significant underestimation of the permeation (see dash lines in Fig. \ref{fig293} and \ref{fig343}).

\subsection{Expectation-Maximization (EM) algorithm}
The pore size distribution of shale matrix is a combination of the pore size distributions within the IOM and OM. The range of pore sizes, however, is different in the two components, with pores in the OM an order of magnitude smaller than those in the IOM\cite{naraghi2015stochastic}. Therfore, the pore structures in the OM and IOM need to be distinguished. Then gas flow with and without surface diffusion of adsorbed gas can be simulated using the GMS and DGM, respectively.

Current nitrogen adsorption tests indicate that the pore size distributions in shale usually satisfy double-mode distributions\cite{al2014comparisons} and the assumption of Gaussian distribution for pore size in rocks has been reported in several studies\cite{naraghi2015stochastic}\cite{kutilek2006influence}. Under the Gaussian mixtures model(GMM) assumption, the Expectation-Maximization (EM) algorithm used by Naraghi and Javadpour\cite{naraghi2015stochastic} is applied here to obtain the pore size distribution in both the IOM and OM. The EM algorithm is an iterative algorithm that starts from the initial estimation of parameters and proceeds to iteratively update those parameters until the convergence is reached. Each iteration consists of an Expectation-step and an Maximization-step. 
\subsubsection{Expectation step}
In the expectation step, the probability that data point $i$ belongs to cluster $j$ can be calculated using the following: 

\begin{equation}
w_j^{(i)} = \frac{g_j(x)\Phi_j}{\sum_{l=1}^k g_l(x)\Phi_l},
\end{equation}
where $w_j^{(i)}$ is the probability that example $i$ belongs to cluster $j$, $k$ is the number of clusters, $\Phi_j$ is the fraction of the dataset belonging to cluster $j$, and $g_j(x)$ is the probability density function of a multivariate Gaussian which satisfies:

\begin{equation}
g_j(x)=\frac{1}{\sqrt{(2\pi)^n|\sum_j|}}e^{\frac{1}{2}(x-\mu_j)^T\sum_j^{-1}(x-\mu_j)},
\end{equation}
where $x$ is the input vector, $n$ is the input vector length, $\sum_j$ is the covariance matrix for cluster $j$, and $\mu_j$ is the mean of cluster $j$. 
\subsubsection{Maximization step}
In the Maximization step, the cluster covariances and means based on the probabilities calculated in the Expectation step are calculated. The updated parameters are expressed in the following equations:
\begin{equation}
\Phi_j^{(new)}=\frac{1}{m}\sum_{i=1}^m w_j^{(i)},
\end{equation}

\begin{equation}
\mu_j^{(new)} = \frac{\sum_{i=1}^m w_j^{(i)}x^{(i)}}{\sum_{i=1}^m w_j^{(i)}},
\end{equation}

\begin{equation}
\sum_j^{(new)} = \frac{\sum_{i=1}^m w_j^{(i)}(x^{(i)}-\mu_j)(x^{(i)}-\mu_j)^T}{\sum_{i=1}^m w_j^{(i)}}.
\end{equation}
\subsubsection{Validation}
\begin{figure}[H]

\centering
\includegraphics[scale=0.27]{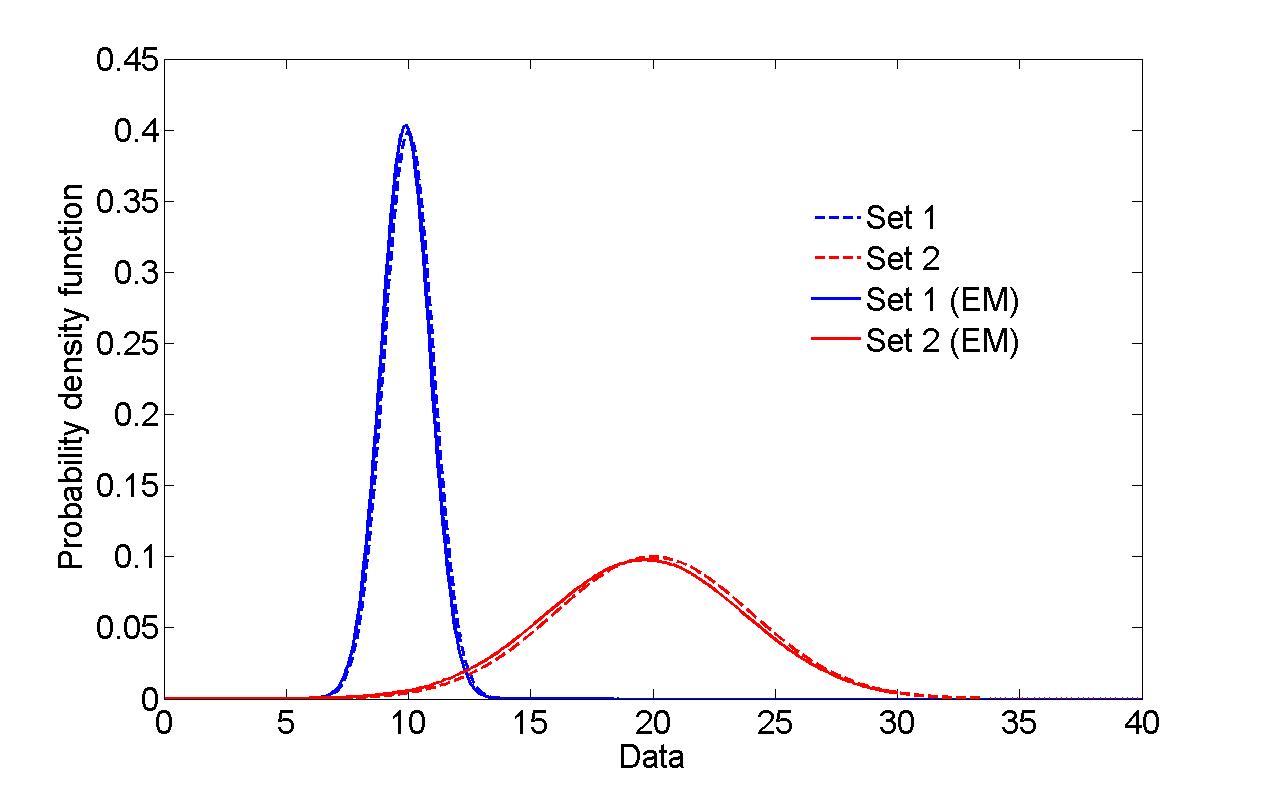}
\caption{Probability density function of two clusters}
\label{figEM}
\end{figure}

To validate the accuracy of the EM algorithm, a two-cluster Gaussian Mixture with different means and variances are tested. The parameters of two clusters and the calculated results are listed in Tab.\ref{tableEM}. The comparison is shown in Fig. \ref{figEM}. The simulation results estimated based on the EM algorithm are comparable to the original data.  
\begin{table}[ht]
\centering
\caption{Parameters of test clusters}
\label{tableEM}
\begin{tabular}{c|c|c|c|c|c|c}
\hline
 	   & $\mu$  & $\sigma$   & Fraction & $\mu_{EM}$  & $\sigma_{EM}$   & $Fraction_{EM}$ \\
\hline
Set 1  &   10   &   1        &  0.25   &9.8848  & 0.9886 &0.2418\\
\hline
Set 2  &    20  &   4        &   0.75  & 19.6846  & 4.0804 & 0.7582\\
\hline

\end{tabular}
\end{table}

\subsection{Generalized lattice Boltzmann model}

Nithiarasu et al.\cite{nithiarasu1997natural} proposed a generalized N-S equation for isothermal incompressible fluid flow in porous media which can be expressed as follows:
\begin{equation}
\nabla\cdot \textbf{u} =0,
\end{equation}

\begin{equation}
\frac{\partial \textbf{u}}{\partial t} + (\textbf{u} \cdot \nabla) \left( \frac{\textbf{u}}{\varepsilon}\right) = -\frac{1}{\rho} \nabla (\varepsilon P) + \upsilon_e \nabla^2 \textbf{u} +\textbf{F},
\end{equation}
where $\textbf{u}$ and $P$ are the volume-averaged velocity and pressure, respectively, $\upsilon_e$ is an effective viscosity equal to the shear viscosity of fluid, $\upsilon$ times the viscosity ratio, and $\textbf{F}$ represents the total body force due to the presence of the porous media and other external forces, which is given by:
\begin{equation}
\label{EqF}
\textbf{F} = -\frac{\varepsilon \upsilon}{K} \textbf{u} - \frac{\varepsilon F_\epsilon}{\sqrt{K}}|\textbf{u}|\textbf{u} + \varepsilon \textbf{G},
\end{equation}
where $\textbf{G}$ is the external force, $F_\epsilon$ is the geometric function and $K$ is the local permeability. To incorporate the effects of Knudsen diffusion and surface diffusion, Eq. \ref{EqF} is modified  by substituting $K$ with apparent permeability $K_{app}$ based on the idea proposed by Chen et al\cite{chen2015generalized}. Without considering the nonlinear drag force term as the flow rate is extremely small in shale matrix, the body force is given by:
\begin{equation}
\textbf{F} = -\frac{\varepsilon \upsilon}{K_{app} \textbf{u}}+\varepsilon\textbf{G},
\end{equation}
and $K_{app}$ can be calculated based on Eq. \ref{Eqkappiom} and Eq. \ref{Eqkappom} for IOM and OM, respectively.

To solve the generalized N-S equation, Guo and Zhao\cite{guo2002lattice} proposed a generalised LBE as follows:
\begin{equation}
f_i(\textbf{x}+\textbf{e}_i \Delta t, t+ \Delta t)- f_i(\textbf{x},t) = -\frac{1}{\tau} [f_i(\textbf{x},t)-f_i^{eq}(\textbf{x},t)] +\Delta t \textit{F}_i,
\end{equation}
where $f_i(\textbf{x},t)$ is the distribution function for the particle with velocity $\textbf{e}_i$  at position $\textbf{x}$ and time t, $\Delta t$ is the time increment, and $\tau$ is the relaxation time. The equilibrium distribution functions $f^{eq}$ are defined as the following form with the porosity effect considered:
\begin{equation}
f_i^{eq} = w_i \rho \left[ 1+\frac{3}{c^2}(\textbf{e}_i \cdot \textbf{u}) + \frac{9}{2\varepsilon c^4} (\textbf{e}_i \cdot \textbf{u})^2 -\frac{3}{2\varepsilon c^2}\textbf{u}^2\right],
\end{equation}
where $w_i$ is the weighting factor and $c=\Delta x /\Delta t$ is the lattice speed.
The force term $\textit{F}_i$ is calculated as:
\begin{equation}
\textit{F}_i = w_i \rho (1-\frac{1}{2\tau})\left[ \frac{3}{c^2}(\textbf{e}_i \cdot \textbf{F}) + \frac{9}{\varepsilon c^4} (\textbf{e}_i \cdot \textbf{u}) (\textbf{e}_i \cdot \textbf{F}) - \frac{3}{c^2}(\textbf{u} \cdot \textbf{F}) \right],
\end{equation}
and the fluid velocity $\textbf{u}$ and density $\rho$ are defined as:
\begin{equation}
 \rho= \sum _i f_i,
\end{equation}

\begin{equation}
\label{equ}
\rho \textbf{u} = \sum_i f_i \textbf{e}_i + \frac{\Delta t}{2}\rho \textbf{F}.
\end{equation}
Due to the quadratic nature of Eq. \ref{equ}, the velocity $\textbf{u}$ can be given explicitly by:
\begin{equation}
\textbf{u}=\frac{\textbf{v}}{c_0+\sqrt{c_0^2+c_1|\textbf{v}|}},
\end{equation}
where $\textbf{v}$ is a temporal velocity defined as:
\begin{equation}
\rho \textbf{v} = \sum_i f_i \textbf{e}_i + \frac{\Delta t}{2}\varepsilon \rho \textbf{G},
\end{equation}
and the two parameters $c_0$ and $c_1$ are given by:
\begin{equation}
c_0 = \frac{1}{2}\left( 1 + \varepsilon \frac{\Delta t}{2} \frac{\upsilon}{K_{app}} \right),
\end{equation}
\begin{equation}
c_1 = \varepsilon \frac{\Delta t}{2} \frac{F_\epsilon}{\sqrt{K_{app}}}.
\end{equation}

The advantages of the above LB model are as follows: First, without invoking any boundary conditions, it can automatically simulate the interfaces between different components in shale matrix with spatially variable porosity and permeability. Second, different nodes in the shale matrix is fully accounted for and the Knudsen diffusion as well as surface diffusion can be easily considered in the model. For more details of the generalized NS equation with slippage effect and the LB model, one can refer to our previous work\cite{chen2015generalized}\cite{chen2015permeability}.

\section{Results and Discussion}

In this section, we first use the GLBM to simulate the $K_{app}$ of both the IOM and OM, and then we use the base case data of an Eagle Ford Shale sample\cite{naraghi2015stochastic} to analyse the $K_{app}$ of reconstructed shale. Different from the previous studies which focus on the influence of porosity, tortuosity and adsorption phenomenon\cite{singh2013nonempirical}\cite{chen2015permeability}\cite{sun2015gas}, the current study details the surface diffusion phenomenon. 

\subsection{Permeability of IOM and OM}

Based on Eq.\ref{Eqkappiom} and Eq.\ref{Eqkappom}, it is evident that $K_{app}$ of the IOM and the OM depend both on pressure and pore radius. $K_{app}$ of the OM also depends on the gas adsorption properties. In this section, the change of $K_{app}$ with $P$ and $r$ both in OM and IOM as well as the variation of $K_{app}$ with Langmuir parameters in OM are discussed. The pore-size distribution is assumed to be Gaussian in both the IOM and OM.

\subsubsection{Permeability of IOM}
\begin{figure}[H]
\centering
\includegraphics[scale=0.23]{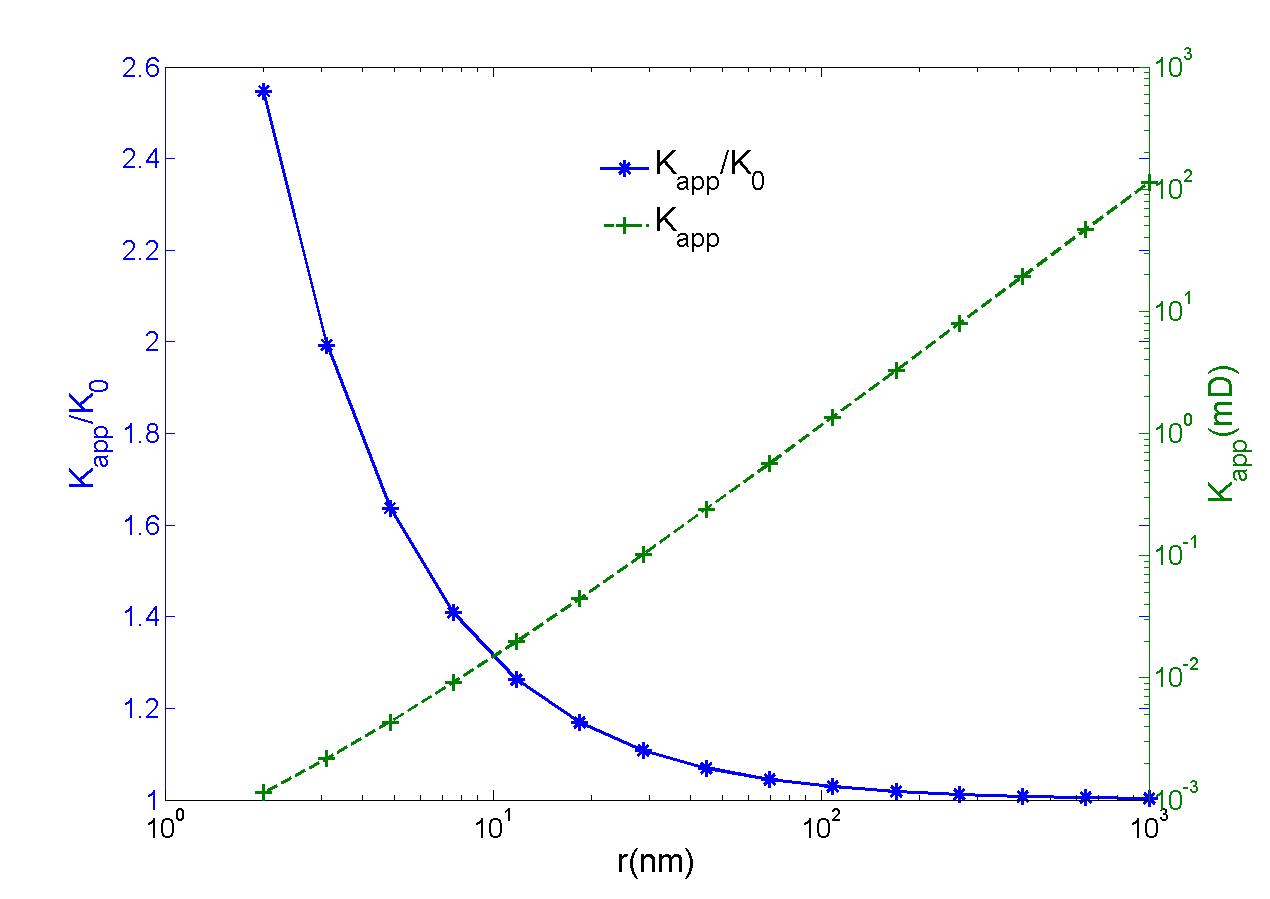}
\caption{Variation of $K_{app}$ and $K_{app}/K_0$ with $r$ (IOM)} 
\label{figiomr}
\end{figure}

Fig.\ref{figiomr} shows the variation of $K_{app}$ and the ratio of $K_{app}$ to intrinsic permeability, $K_0$ with the mean pore size distribution of IOM, and the input parameters are listed in Tab.\ref{tabIOM}. Based on Fig.\ref{figiomr}, one can observe that with the increase of the mean pore size, $K_{app}$ increases, so does its rate of change. $K_{app}/K_0$ and its slope, however, decrease with the increase of mean pore size. When the mean pore size approaches $100 nm$, the gas flow becomes continuum, and $K_{app}$ approximately equals $K_0$. The simulation results clearly show the effect of Knudsen diffusion on the gas transport in IOM, especially when pore radius is smaller than 100nm. Also the deviation of $K_{app}$ from $K_0$ indicates that the permeability of IOM in shale will be underestimated if the Knudsen diffusion is ignored. 
\begin{table}[ht]
\centering
\caption{Input parameters of IOM}
\label{tabIOM}
\begin{tabular}{c|c}
\hline
Inlet pressure $P_{in}$ ($MPa$) & 22 \\
\hline
Outlet pressure $P_{out}$ ($MPa$) &20 \\
\hline
Viscosity of Methane at outlet ($10^{-6}Pa \cdot s$)  &18.231  \\
\hline
Tortuosity of IOM  & 2 \\
\hline
Porosity  of IOM   & 0.04\\
\hline
Length $\times$ Width ($cm\times cm$) & 3.2 $\times$ 1.6 \\
\hline
\end{tabular}
\end{table}

\begin{figure}[H]
\centering
\includegraphics[scale=0.23]{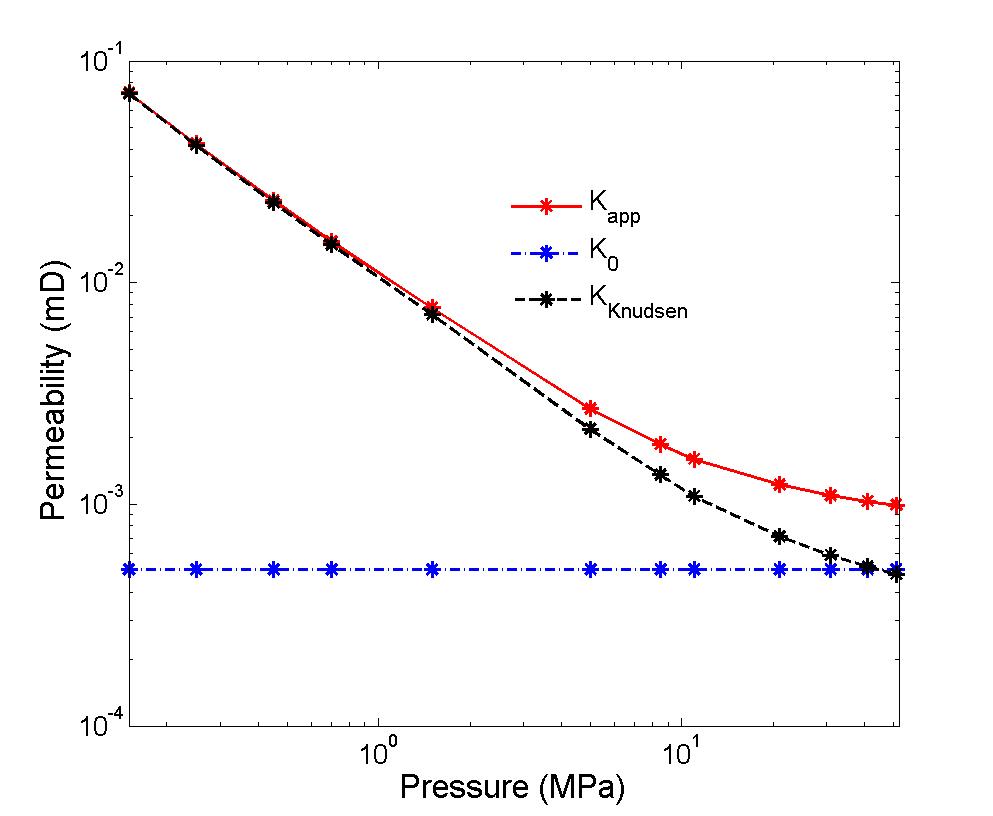}
\caption{Variation of $K_{app}$ and its components with average pressure(IOM)}
\label{figiomp}
\end{figure}

Fig.\ref{figiomp} shows the variation of $K_{app}$, $K_0$, and $K_{Knudsen}$ with the average pressure, where $K_0$ and $K_{Knudsen}$ are components from viscous flow (intrinsic permeability) and Knudsen diffusion, respectively. The average radius is 2 $nm$ and the variance is 0.25. As can be seen in Fig.\ref{figiomp}, as the average pressure increases, the intrinsic permeability does not change, but $K_{Knudsen}$ (and hence $K_{app}$) decreases. This result is important for shale gas production, as the reservoir is depleted and the reservoir pressure decreases, $K_{app}$ deviates from the intrinsic permeability owing to high contribution of Knudsen diffusion at zones with low pressures, and gas transport in these zones therefore is predominantly controlled by Knudsen diffusion. Fig.\ref{figiomp} also indicates that permeability of IOM is a function of reservoir pressure, and the permeability must be considered as a dynamic reservoir parameter and updated accordingly as the reservoir is being depleted. 

Similar dependencies on pore sizes and pressure have been observed by authors considering conductivity of a single channel\cite{javadpour2009nanopores} and bundle of channels or nanopore networks\cite{sakhaee2012gas}.

\subsubsection{Permeability of OM}

Adsorbed gas and surface diffusion in OM are drawing more attention recently. Various studies have confirmed that the adsorption can change $K_{app}$ of shale matrix and the surface diffusion is an important transport mechanism at reservoir conditions\cite{singh2013nonempirical}\cite{do1998dual}. In some treatments, the surface diffusion is considered as an extra flux in capillary models\cite{ren2015lattice}\cite{xiong2012fully}\cite{sun2015gas}. However the sensitivity of permeability to adsorption and surface diffusion are seldom mentioned. In others, the adsorbed gas is modelled by taking some pore space and therefore reducing the local pore size\cite{chen2015permeability}\cite{allan2013effect}. In other words, if adsorbed gas is assumed to pre-exists in OM, the intrinsic permeability will increase with the decrease in pressure because the desorbed gas will release some pore space\cite{singh2013nonempirical}. Nonetheless, no consensus on the volume occupied by adsorbed gas exists in these treatments\cite{xiong2012fully}\cite{allan2013effect}\cite{sun2015gas}, also the contribution of the surface diffusion is ignored. In this study, with the introduction of DGM-GMS model, the effect of surface diffusion on $K_{app}$ can be easily considered, moreover, based on the assumption that the total flux $N_{total}$ is the sum of $N_{DGM}$ and $N_{GMS}$, the ambiguity in the adsorption layer thickness can be avoided. As mentioned above, the contribution of surface diffusion is affected by parameters regarding the Langmuir adsorption model and the surface diffusivity. Therefore, in this study, all model parameter values were carefully collected from the reported data. Otherwise, unreasonable values may lead to misleading conclusions. 

\begin{figure}[H]
\centering
\includegraphics[scale=0.42]{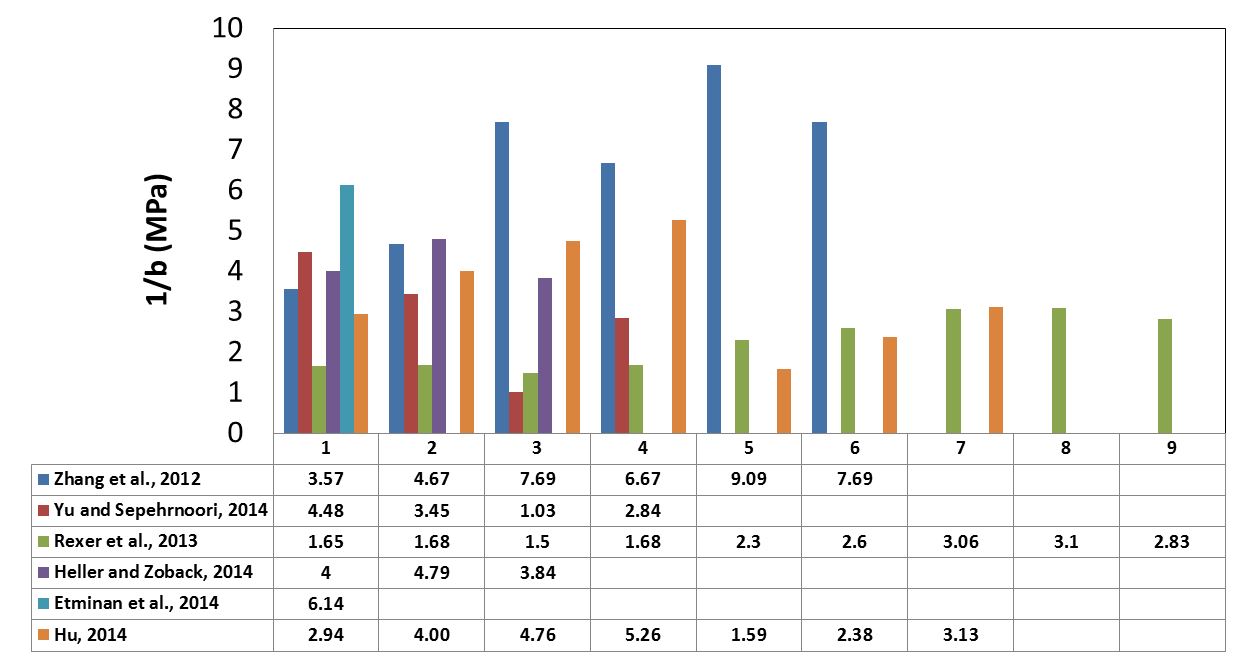}
\caption{Langmuir constant given by different authors\cite{zhang2012effect}\cite{yu2014simulation}\cite{rexer2013methane}\cite{heller2014adsorption}\cite{etminan2014measurement}\cite{hu2014methane}}
\label{figb}
\end{figure}

$b$ and $q_{sat}$ can be collected from the Langmuir isothermal of shale samples. Fig. \ref{figb} summarises current experimental data for $b$ for different shale samples. Based on these data, $b$ approximately varies from $0.125$ to $1 MPa^{-1}$($b^{-1}$ from $1 $ to $ 8 MPa$).    

\begin{figure}[H]
\centering
\includegraphics[scale=0.42]{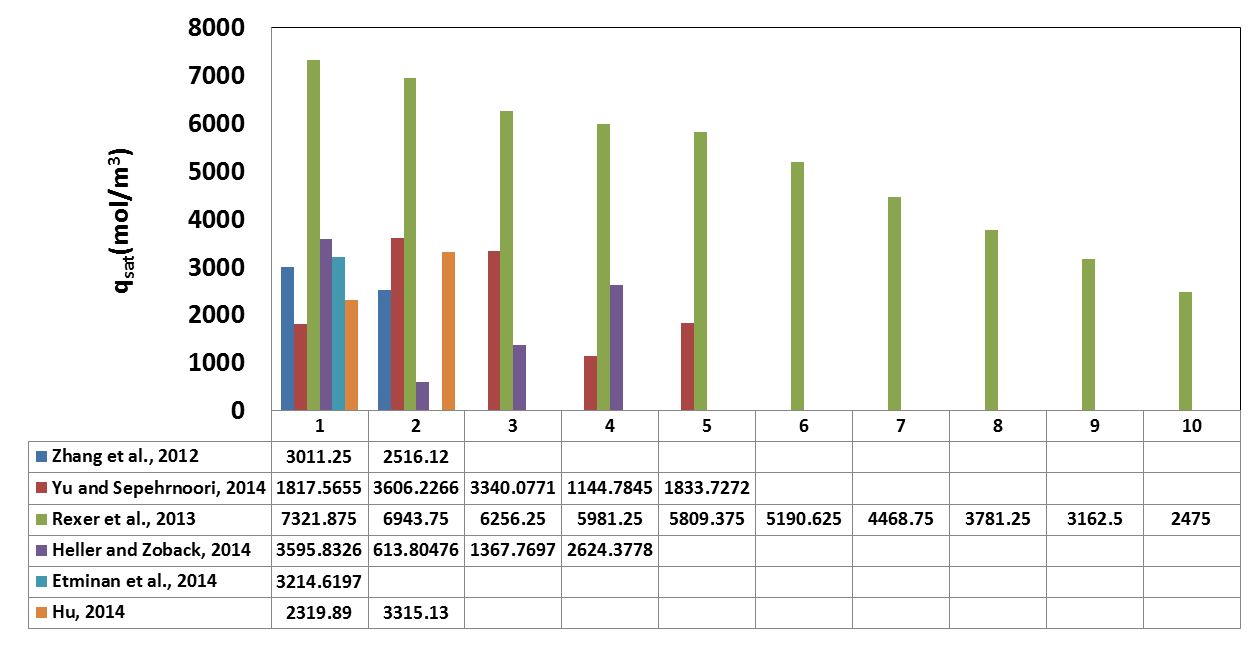}
\caption{Langmuir volume given by different authors\cite{zhang2012effect}\cite{yu2014simulation}\cite{rexer2013methane}\cite{heller2014adsorption}\cite{etminan2014measurement}\cite{hu2014methane}}
\label{figqsat}
\end{figure}

For the determination of Langmuir volume $q_{sat}$, data should be treated with caution as the definition and unit of $q_{sat}$ vary from one study to another. For instance, Eq.\ref{EQqsat} is used to obtain $q_{sat,k}$ with a unit of $``mol/m^3" $ if $q_{sat,b}$ is defined as `` $m^3/kg $ ".  

\begin{equation}
\label{EQqsat}
q_{sat,k}=\frac{q_{sat,b} \rho_b}{(1-\varepsilon)  TOC} \frac{\rho_{CH_4}}{M_{CH_4}},
\end{equation}
where TOC is the volume fraction of the kerogen content. 
$q_{sat,k}$ is the mole volume of methane divided by the volume of kerogen skeleton. $q_{sat,b} = V/m_b$ is the volume of methane divided by the bulk mass of shale sample.  
With the assumption that the porosity and the density of kerogen is 0.2 and $1.65 \times 10^3 kg/m^3$, $q_{sat,k}$ approximately ranges from $1000$ to $8000 mol/m^3$(see Fig. \ref{figqsat}). 

The surface diffusion diffusivity can be measured in forms of self-diffusivity $D_{self}$, Maxwell-Stefan diffusivity $D_s$ or Fick diffusivity $D_f$. From Tab.\ref{tabcoefficient} it can be seen that $D_f > D_s > D_{self}$. The $D_f$ of organic shale given by Kang et al.\cite{kang2011carbon} is from $1.55 \times 10^{-7}$ to $8.8 \times 10^{-6} m^2/s$, and by Akkutlu and Fathi\cite{akkutlu2012multiscale} is from $8.3 \times 10^{-8}$ to $8.8 \times 10^{-6} m^2/s$. $D_{self}$ of organic shale given by Zhai et al.\cite{zhai2014adsorption} is between $9.43 \times 10^{-9}$ and $11.62 \times 10^{-9} m^2/s$ and by Yuan et al.\cite{yuan2014experimental} is from $2.38 \times 10^{-9}$ to $9.96 \times 10^{-9} m^2/s$.  Wu et al.\cite{wu2015model} proposed an empirical equation to determine $D_s$ of methane based on the experimental data of methane-activated carbon:

\begin{equation}
D_s = (8.29 \times 10^{-7})T^{0.5}exp\left(-\frac{\Delta H^{0.8}}{RT}\right),
\end{equation}
where $\Delta H$ is the isosteric adsorption heat, ranging from $12000$ to $16000 J/mol$ at $338.7K$, and $D_s$ ranges from $6.72 \times 10^{-6}$ to $7.96 \times 10^{-6} m^2/s$. Based on these data, a reasonable range of Maxwell-Stefan diffusivity used in this study is set from $1\times 10^{-9}$ to $1 \times 10^{-5} m^2/s$.
\begin{table}[ht]
\centering
\caption{Comparison of different diffusivity\cite{ramanan2006modeling}}
\label{tabcoefficient}
\begin{tabular}{c|c}
\hline
Self-diffusion diffusivity & $D_{self} \approx D_0 (1-\theta)$  \\
\hline
Maxwell-Stefan diffusivity & $D_s \approx D_0$  \\
\hline
Fick diffusion diffusivity  &$D_f = \frac{D_0}{1-\theta}$  \\
\hline
\end{tabular}
\end{table}

\begin{figure}[H]
\centering
\includegraphics[scale=0.25]{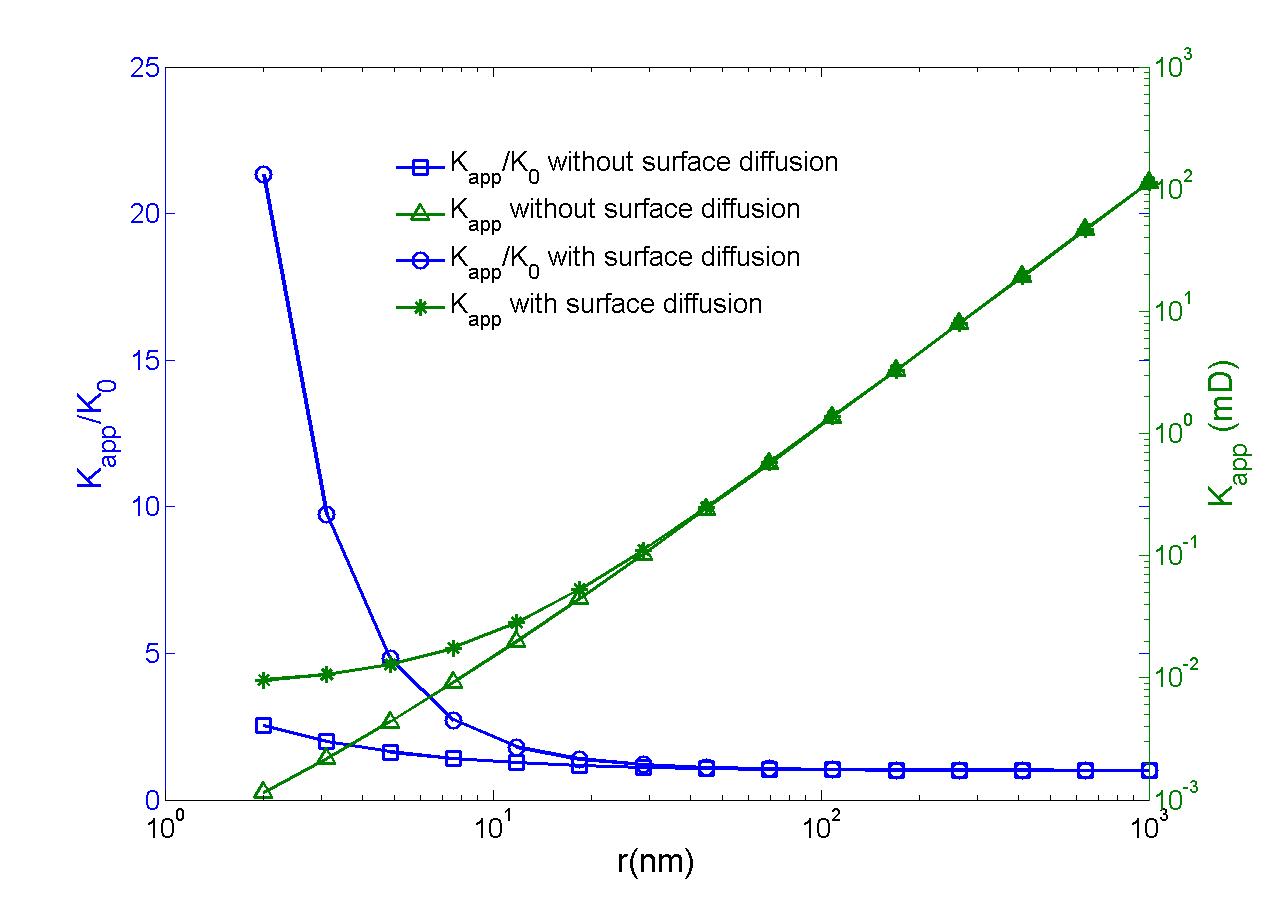}
\caption{Variation of $K_{app}$ and $K_{app}/K_0$ with r(OM)}
\label{figom_r}
\end{figure}

Fig.\ref{figom_r} shows the variation of $K_{app}$ and $K_{app}/K_{0}$ of OM with $r$. $D_s$,  $q$ and $b$ are set to be $5\times 10^{-7} m^2/s $, $4000 mol/m^3$ and $0.25 MPa^{-1}$, respectively. The other inputs are the same as that listed in Tab. \ref{tabIOM}. $K_{app}$ without considering the surface diffusion is also listed in the figure for comparison. From Fig.\ref{figom_r} it can been seen that, the trend of $k_{app}$ and $k_{app}/k_0$ of OM is similar to that of IOM. In large pores, both Knudsen diffusion and surface diffusion can be ignored, and $K_{app}$  equals $K_0$. Fig.\ref{figom_r} also shows that the surface diffusion is the primary flow mechanism in nanopores, and significant underestimation of permeability can be observed if the surface diffusion is ignored when the average pore radius is less than 10 nm. Wu et al.\cite{wu2015model} reported that the contribution of surface diffusion to the shale gas mass transfer can be up to $92.95\%$ when pore radius is smaller than 2$nm$. The results presented in Fig.\ref{figom_r} agree well with that presented by Wu et al.\cite{wu2015model}study, and the value of $K_{app}/K_{0}$ can be larger than 10 when $r < 2nm $.

\begin{figure}[H]
\centering
\includegraphics[scale=0.3]{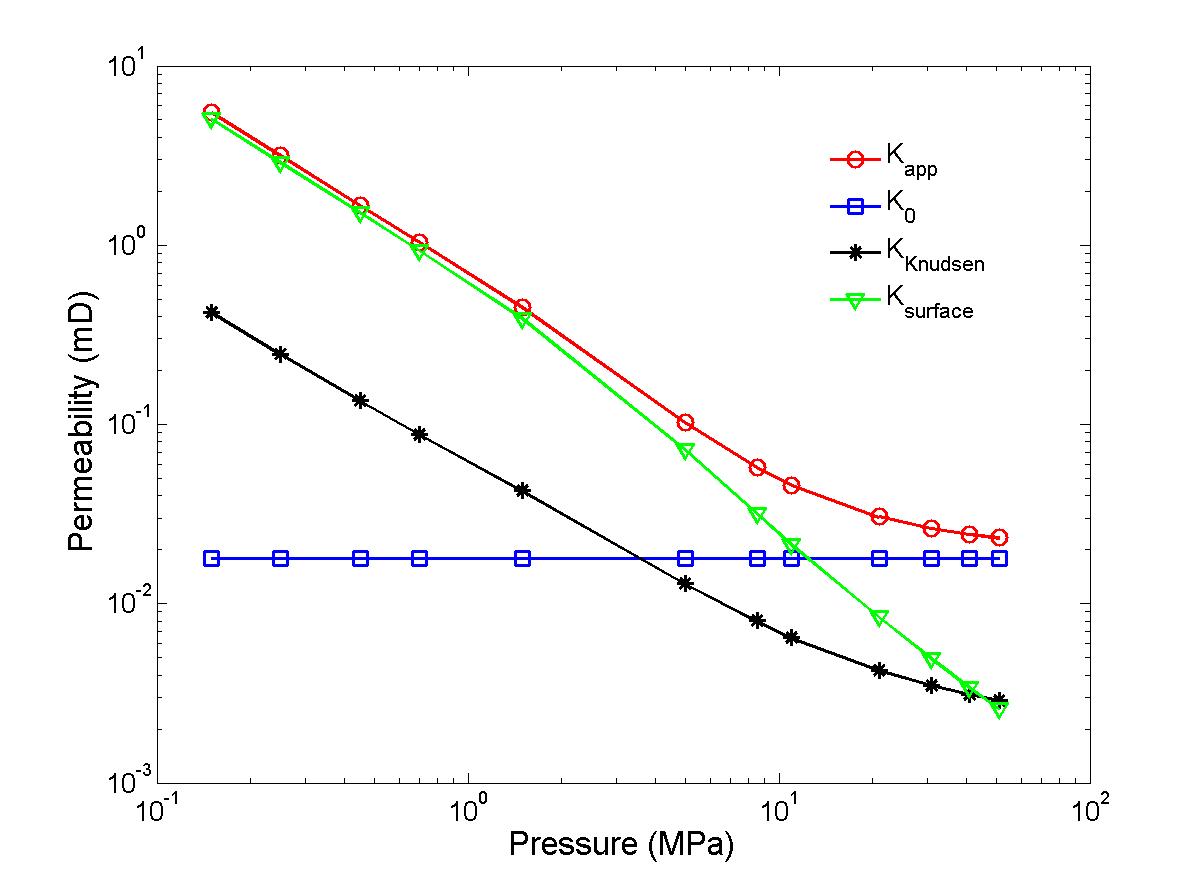}
\caption{Variation of $K_{app}$ and its components with average pressure(OM)}
\label{figom_p}
\end{figure}

Fig.\ref{figom_p} illustrates $K_{app}$ of OM estimated based on DGM-GMS under different average pressure conditions. The average radius used in this case is 2 $nm$ with a variance of 0.25. The other inputs are the same as those of Fig.\ref{figom_r}. In Fig.\ref{figom_p}, the surface diffusion decreases linearly with the increasing pressure. The reason is that the physical sorption force increases with pressure, and therefore the high pressure restricts the movement of adsorbed gaseous molecules. When the pressure reaches a high level, the contribution from both Knudsen diffusion and surface diffusion becomes almost negligible. The surface diffusion, however, is the dominant mass transport mechanism at low pressures, and $K_{app}$ accounting for surface diffusion can be several times larger than that without considering it. 

\begin{figure}[H]
\centering
\includegraphics[scale=0.3]{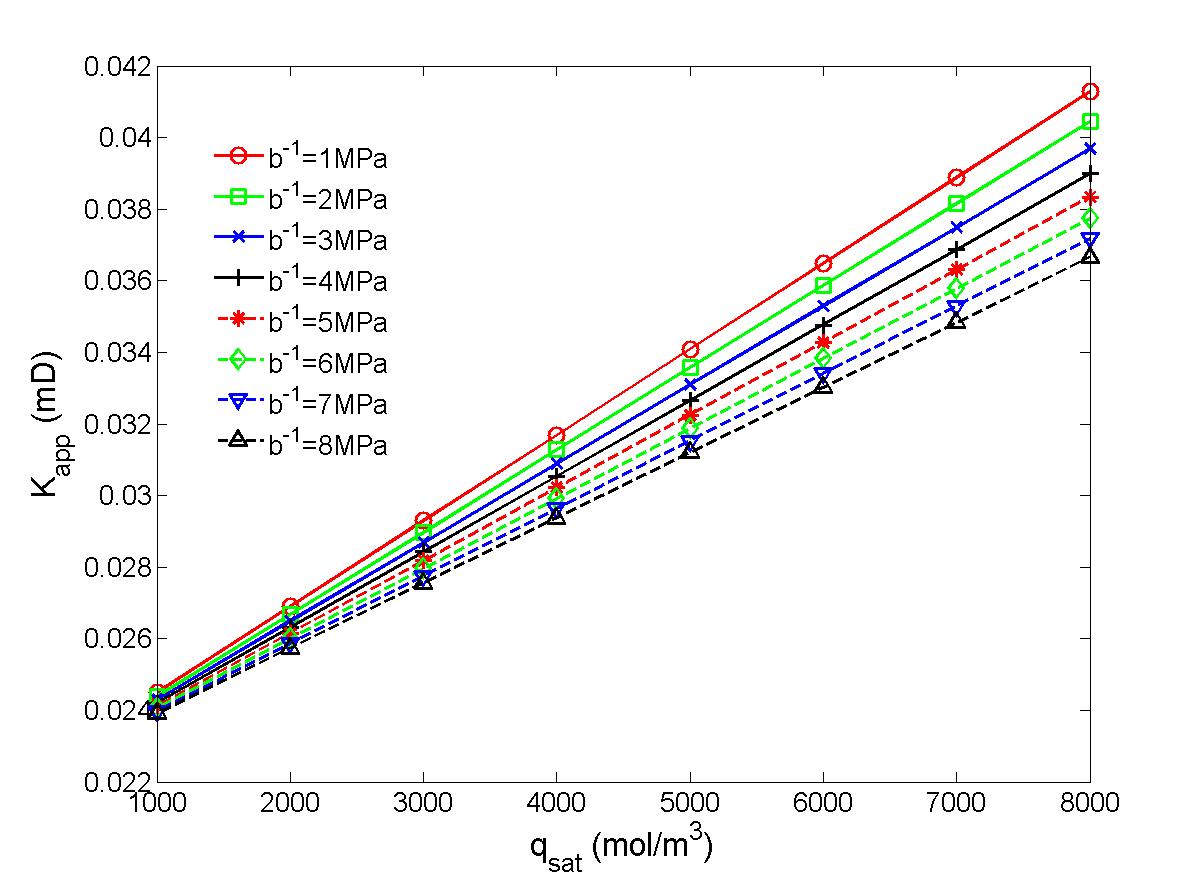}
\caption{Variation of $K_{app}$ with Langmuir parameters}
\label{figom_Langmuir}
\end{figure}

Fig.\ref{figom_Langmuir} depicts $K_{app}$ of OM with different Langmuir parameters. $D_s$ is set to be $5\times 10^{-7} m^2/s$, the average radius is 11 $nm$, and $\sigma$ is 0.25. To represent the reservoir condition, the inlet pressure is set to be $22 MPa$, and the outlet pressure is $20 MPa$. As can be seen in Fig.\ref{figom_Langmuir}, $K_{app}$ is linearly increasing with the increase of $q_{sat}$ and the decrease of $b^{-1}$. This is because based on Langmuir model(Eq.\ref{EqsimpleLangmuir}), a larger $q_{sat}$ or $b$ induces a larger amount of adsorbed gas at a certain pressure, and therefore leads to a higher surface flux. Also, the influence of $b$ on $K_{app}$ is becoming obvious with the increase of $q_{sat}$. The results of Fig.\ref{figom_Langmuir} further emphasizes the impact of adsorbed gas on gas flow in OM.    

The above results confirm that the flow patterns of OM differ from that of IOM, and the gas flow in OM is more complicated than that in IOM. Moreover, the variation of $K_{app}$ with pore size distribution suggests that the pore sizes corresponding to each components should be considered, and the usage of the mean pore size could be erroneous. In the following sections, the analysis based on reconstructed shale matrix consisting of both OM and IOM are performed to estimate the variation of $K_{app}$ with component distribution and content. 

\subsection{Effects of the component distribution}
\begin{figure}[H]
\centering
\includegraphics[scale=0.5]{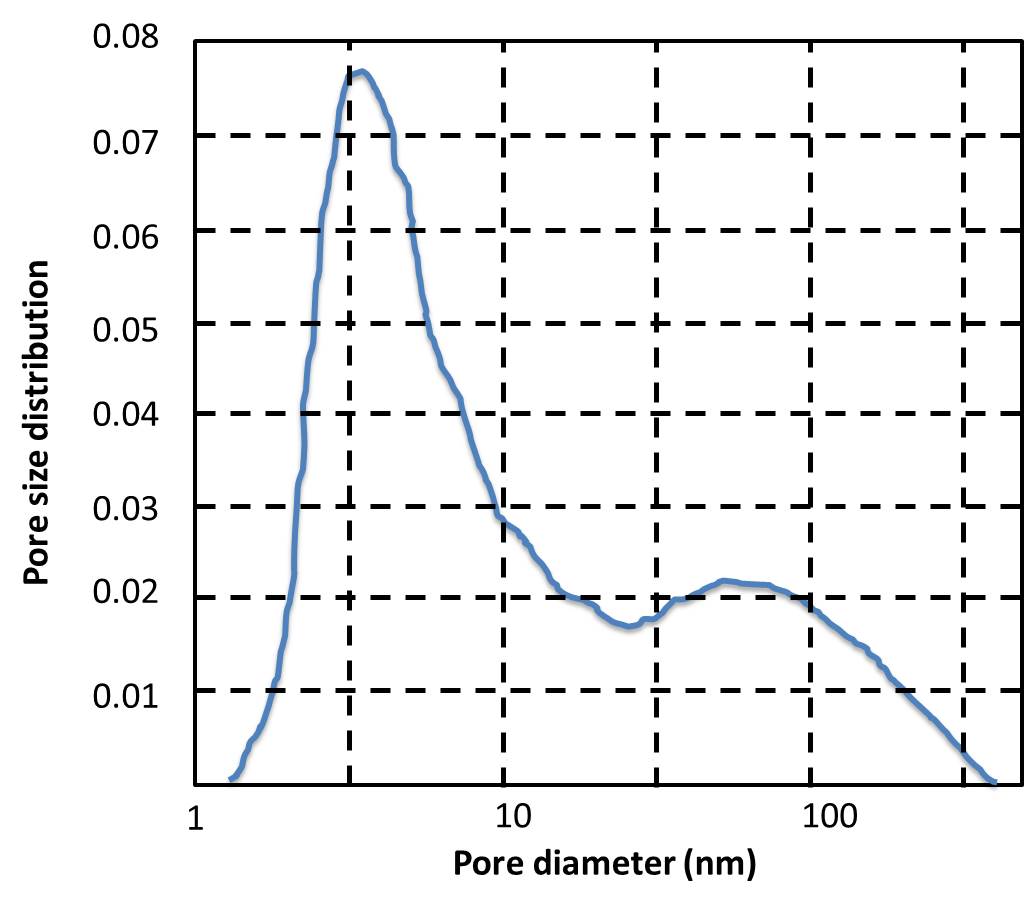}
\caption{Pore-size distribution of the Eagle Ford Shale sample}
\label{figpsd}
\end{figure}

The pore size distribution of reconstructed shale sample is obtained by nitrogen intrusion-test of the Eagle Ford Shale reported by Naraghi and Javadpour\cite{naraghi2015stochastic}. As can be seen in Fig.\ref{figpsd}, an evidenced bimodal trend exists in the shale sample. With a hypothesis that the pore radius of OM is smaller than that of IOM, pore size distributions in OM and IOM are extracted from Fig.\ref{figpsd} by using EM algorithm and are presented in Tab.\ref{tabpsd}.
 
\begin{table}[ht]
\centering
\caption{Properties of pore size distributions in OM and IOM}
\label{tabpsd}
\begin{tabular}{c|c|c|c}
\hline
Region  & $\mu ($nm$)$  &$\sigma$   & Fraction \\
\hline
OM      & 0.4    &0.18       & 0.56  \\
\hline
IOM     &1.4     &0.44       &0.44    \\
\hline
\end{tabular}
\end{table}

\begin{figure}[H]
\centering
\includegraphics[scale=0.41]{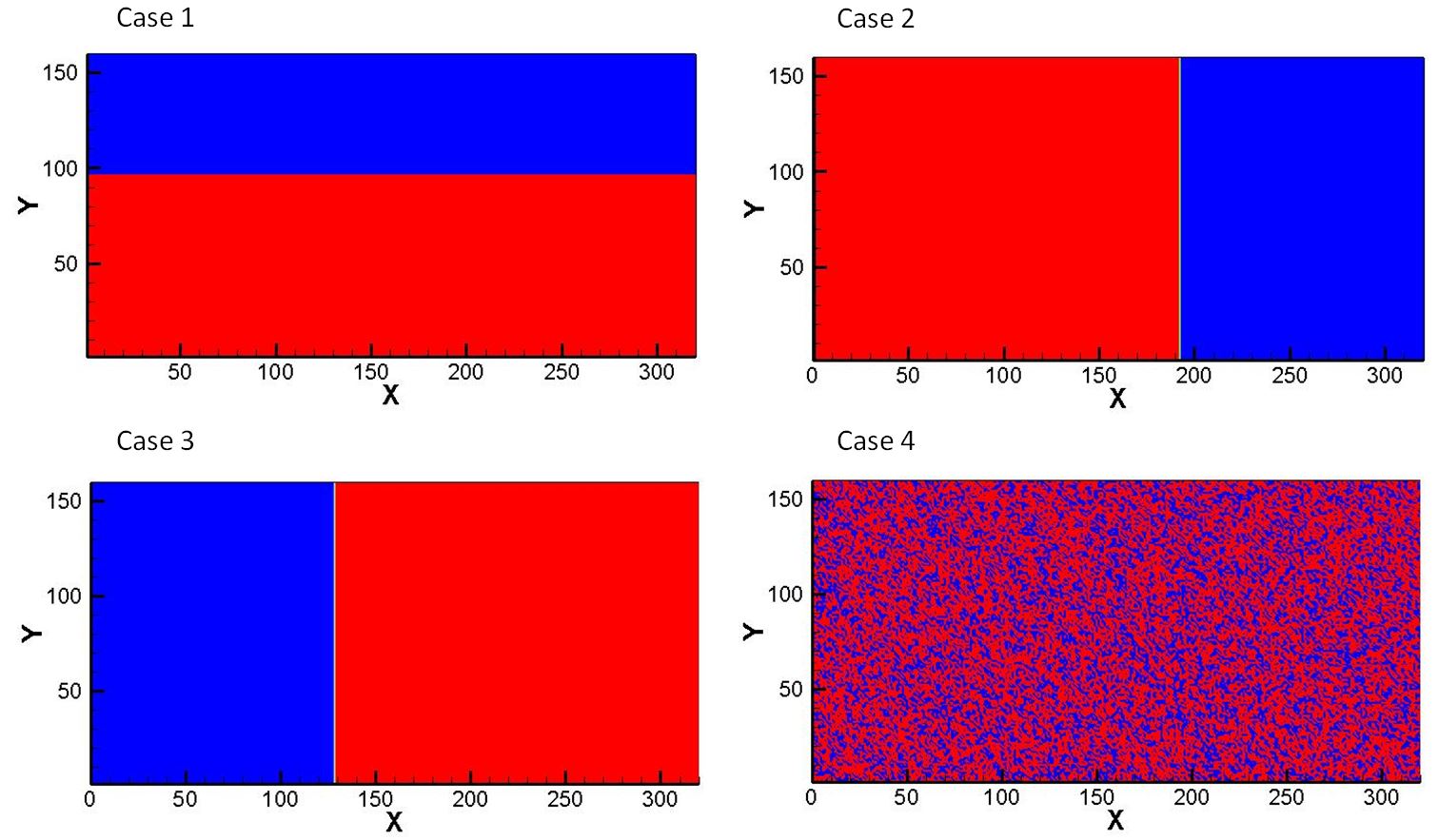}
\caption{Structure of reconstructed shale matrix(the region in red is OM, and the region in blue is IOM)}
\label{figshale_structure}
\end{figure}

Based on the model parameters, the pore size distribution in different OM and IOM regions can be found. In order to study the effect of the distribution of each component on $K_{app}$, four cases are considered( See Fig.\ref{figshale_structure} ). In case 1, the OM and the IOM are placed parallel to the flow direction. In case 2 and case 3, gas passes through the OM and the IOM in sequence. In case 4, the OM and the IOM are randomly distributed in the simulation domain. For each case, multiple realizations are generated randomly into different data sets based on the parameters listed in Tab.\ref{tabpsd}. The other inputs are listed in Tab.\ref{tabinput_shale}.

\begin{table}[ht!]
\centering
\caption{Input parameter for gas flow in shale matrix}
\label{tabinput_shale}
\begin{tabular}{c|c}
\hline
$P_{in}$($MPa$) &0.2 \\
\hline
$P_{out}$($MPa$) &0.1 \\
\hline
$q_{sat}$ ($mol/m^3$) &4000 \\
\hline
$D_s$ ($m^2/s$)   & $1\times10^{-8}$\\
\hline
$b$ ($MPa^{-1}$)  &0.25 \\
\hline
Tortuosity    &2\\
\hline
Porosity of IOM &0.02\\
\hline
Porosity of OM  &0.04\\
\hline
Length $\times$ Width ($\mu m \times \mu m$) & 16 $\times$ 8 \\
\hline
\end{tabular}
\end{table}

\begin{figure}[H]
\centering
\includegraphics[scale=0.3]{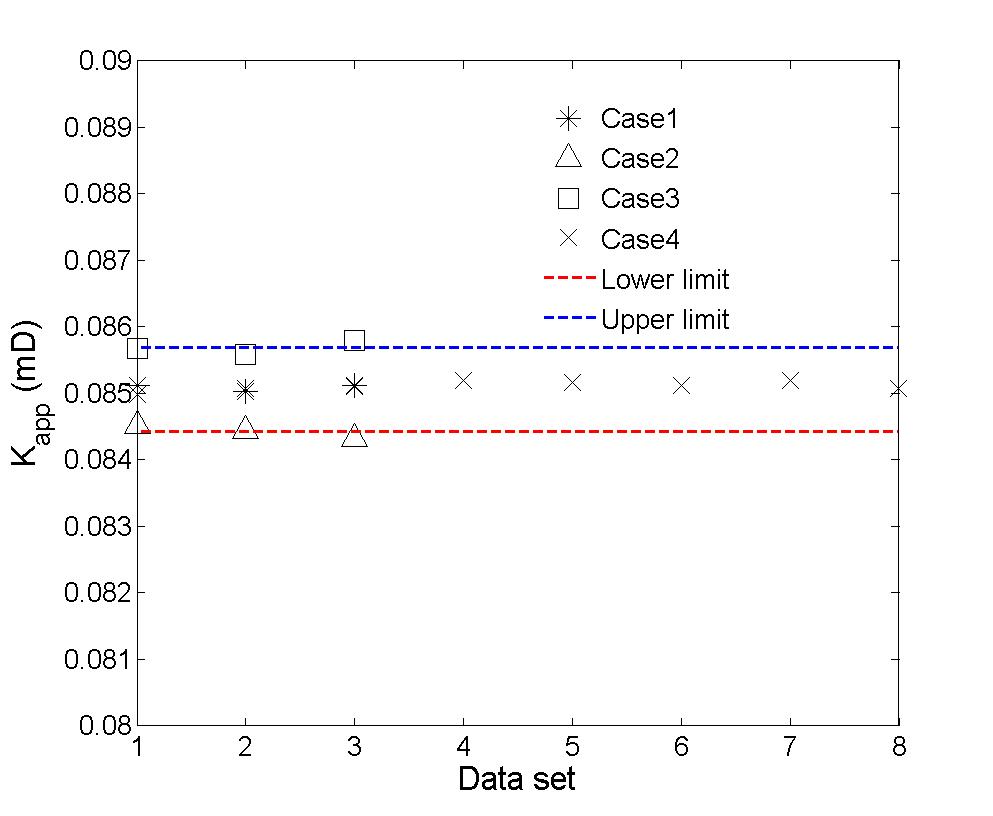}
\caption{$K_{app}$ of shale matrix with different component distribution}
\label{figkapp_shale}
\end{figure}

The simulation results of $K_{app}$ for different component distributions are listed in Fig.\ref{figkapp_shale}. It can be be seen that for each case, different data sets give rise to almost identical results, which demonstrates the numerical accuracy of the current LB model. Comparing all four cases, it can be seen that the influence of the component distribution on $K_{app}$ is not significant, and $K_{app}$ approximately ranges from 0.0845 $mD$ to 0.0856 $mD$. Nevertheless, within this small range of $K_{app}$, case 3 and case 2 determine the upper and lower bounds of $K_{app}$, respectively.   It has been pointed out that gas becomes more rarefied in smaller pores under lower pressure. As the mean pore radius of OM is roughly 3 times smaller than that of IOM, accumulation of OM near the outlet (case 3) where the pressure is lower will increase the gas rarefaction, leading to a higher $K_{app}$. Conversely, when OM is accumulating near the inlet (case 2), the gas rarefaction in OM pores is less significant due to the high pressure.  

\subsection{Effects of  the total organic content (TOC)}

\begin{figure}[H]
\centering
\includegraphics[scale=0.41]{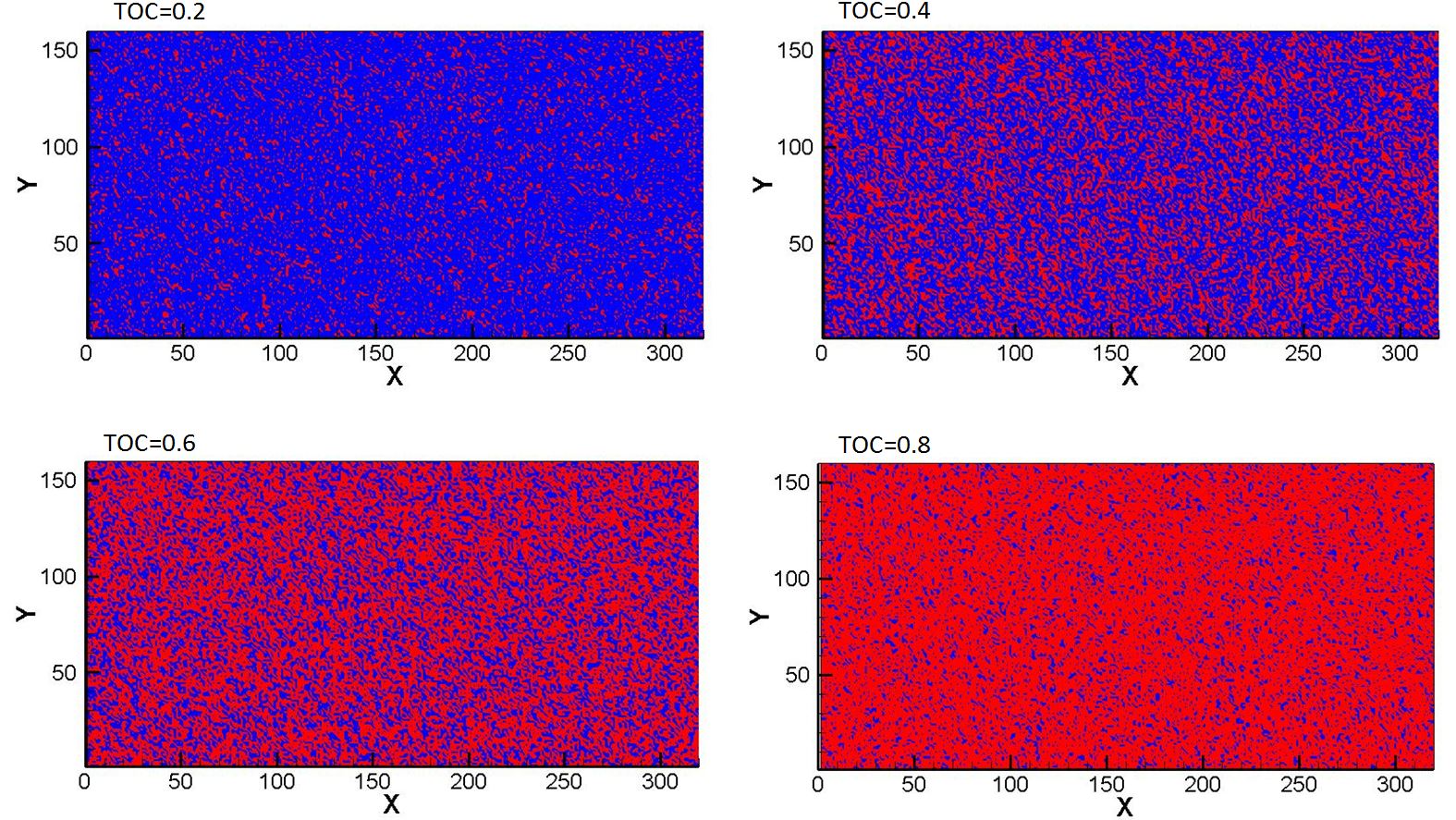}
\caption{The structure of shale matrix with different TOC (The region in red is OM, and the region in Blue is IOM)}
\label{figtoc}
\end{figure}

To study the effects of TOC on permeability of shale, the shale matrix is constructed based on the pore size distributions listed in Tab.\ref{tabpsd}  with the TOC varying from 0.2 to 0.8. The structures of the shale matrix are shown in Fig.\ref{figtoc}, and the pore size distributions within both OM and IOM are assumed to remain constant during the simulation which are the same as that listed in Tab.\ref{tabinput_shale}. 
 
\begin{figure}[H]
\centering
\includegraphics[scale=0.28]{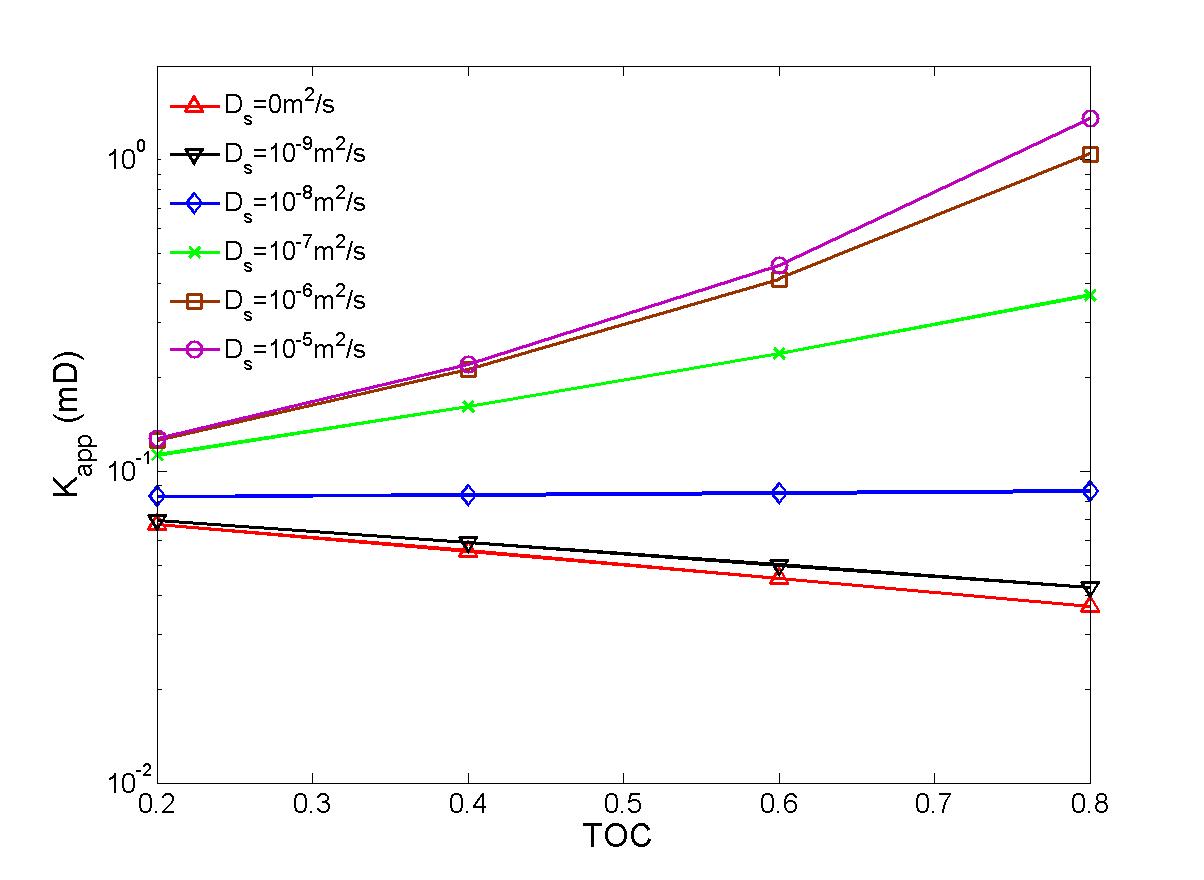}
\caption{Variation of $K_{app}$ with TOC}
\label{figkwithtoc}
\end{figure}

As have been discussed in section 3,  $K_{app}$ is influenced by the Langmuir adsorption properties and the surface diffusivity, therefore the variation of $K_{app}$ with different $D_s$ are analysed and the simulation results are presented in Fig.\ref{figkwithtoc}. When the surface diffusion is negligible, $K_{app}$ decreases with TOC as OM has smaller pores. A similar phenomenon has been observed by Naraghi and Javadpour\cite{naraghi2015stochastic} based on their stochastic permeability model for shale gas systems. With the increase of $D_s$, the contribution of surface diffusion to total flux is becoming more pronounced, and therefore the negative impact of small pore radius on $K_{app}$ is gradually compensated. In current simulation study, when $D_s$ is larger than $1 \times 10^{-8} m^2/s$, the permeability starts to increase with TOC. 

\begin{figure}[H]
\centering
\includegraphics[scale=0.43]{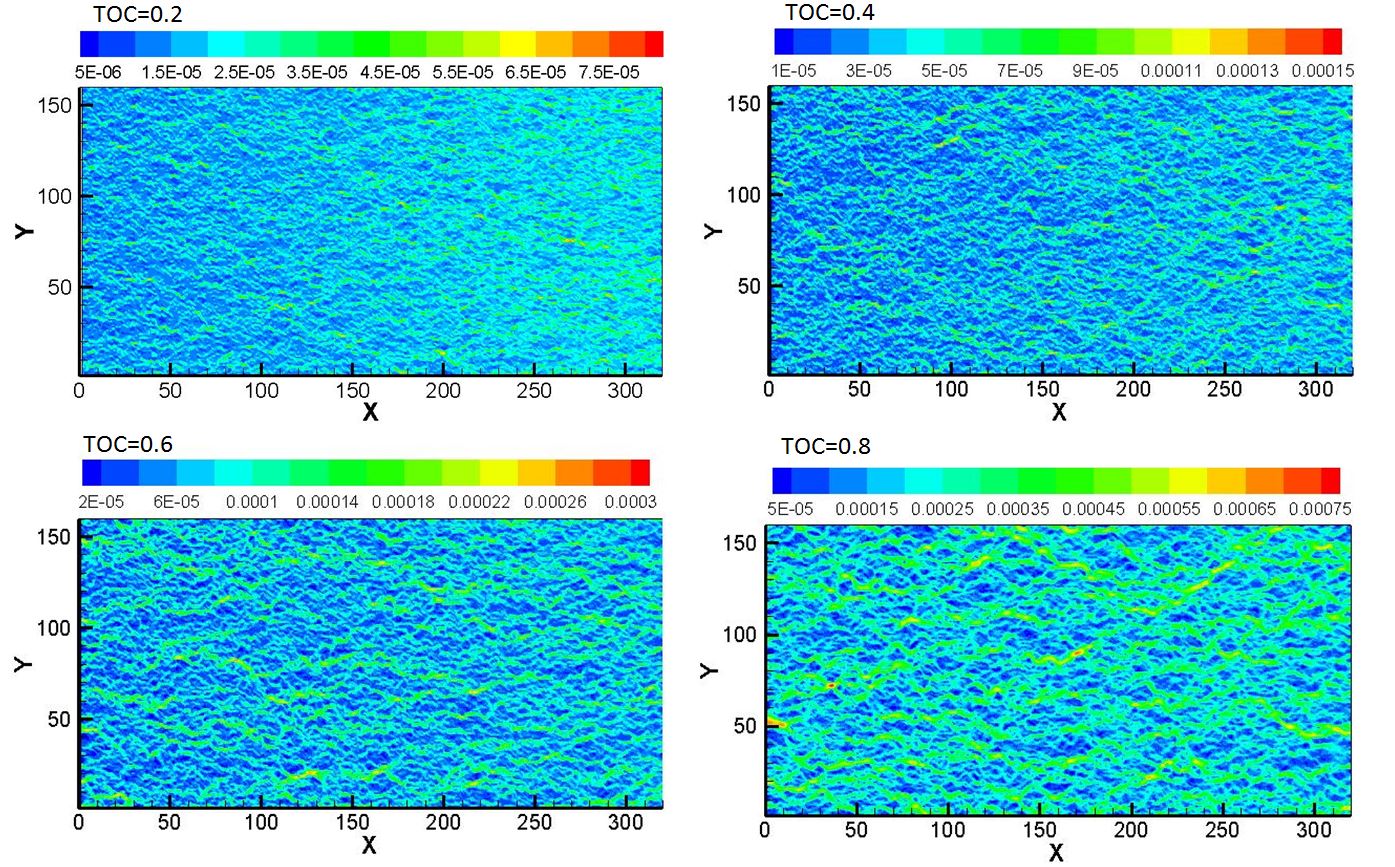}
\caption{Velocity distribution when $D_s = 10^{-5} m^2/s$}
\label{figv_distribution}
\end{figure}

The velocity distributions in the reconstructed shale matrix with different TOC are shown in Fig.\ref{figv_distribution}. In this study, $D_s$ is equal to $1 \times 10^{-5} m^2/s$. As the surface diffusion of adsorbed gas provides an extra mass flux in addition to the Knudsen diffusion and the convection flow, at this high surface diffusivity, the velocity magnitude increases with the increase of the TOC. When TOC=0.2, the velocity ranges from $5\times 10^{-6}$ to $7.5\times 10^{-5} m/s$. However, the velocity is 10 times larger when TOC=0.8. Moreover, from Fig.\ref{figv_distribution} it can been seen that, some predominate pathways are formed with the increase of TOC, which again emphasizes the contribution of surface diffusion to gas flow in OM. 

\section{Conclusion}
In this study, a generalised lattice Boltzmann algorithm for gas flow simulation through porous shale matrix and the DGM-GMS model are employed to estimate $K_{app}$ that includes the surface diffusion effect. Numerical simulations of gas flow in reconstructed shale matrix containing the organic matter and inorganic matter have been carried out.   

Unlike the intrinsic permeability, $K_{app}$ of the shale matrix is a dynamic parameter and should be adjusted as the pressure decreases in a reservoir. Permeability of the IOM increases with the increase of pore radius and the decrease of pressure. Permeability of the OM shows a similar trend with pressure and pore radius to that of the IOM. Moreover, surface diffusion in the OM can have a more important role than Knudsen diffusion and convection flow in determining the apparent permeability at low pressure and in pores smaller than $10 nm$. Furthermore, permeability shows strong dependence on the value of Langmuir parameters and surface diffusivity, which is still not well understood. 

Simulation results based on the reconstructed shale matrix show that the distribution of the OM and the IOM has negligible influence on $K_{app}$ of organic shale. The TOC, on the other hand, plays a significant role in the determination of $K_{app}$, and its effect depends on the surface diffusivity of adsorbed gas.    

The present study emphasizes that the surface diffusion in OM needs to be considered for an accurate determination of the apparent permeability, especially for very small pores at low pressure. The proposed LBM is shown to be an effective simulation tool in determining the apparent permeability of organic shale and in revealing gas transport mechanisms in shale matrix. In future studies, $K_{app}$ based on the real shale sample will be analysed.
\section{Acknowledgement}
The authors would like to acknowledge the support from SCOPE, UNSW and the LDRD program of LANL. J.W. would like to acknowledge the financial support from the China Scholarship Council(CSC). J.W. also thanks the suggestions from Prof. Andreas Seidel-Morgenstern, Max Planck Institute for Dynamics of Complex Technical Systems, Magdeburg. L.C. would also like to acknowledge the support from National Nature Science Foundation of China(No. 51406145,51136004), and Q.K. would also like to acknowledge the support from a DOE oil $\&$ gas project.
\bibliographystyle{unsrt}
\bibliography{mybibfile}
\end{document}